\pdfoutput=1
\documentclass[a4paper]{cas-dc}

\usepackage[sort,compress,numbers]{natbib}
\usepackage{booktabs}
\usepackage{hyperref}
\usepackage{multirow}
\usepackage{graphicx}
\usepackage{tabularx}
\usepackage{siunitx}
\usepackage{float}

\begin{document}
\shortauthors{H.~Ma et al.}
\shorttitle{In-situ measurements in CJPL-II}
\title[mode = title]{In-situ gamma-ray background measurements for next generation CDEX experiment in the China Jinping Underground Laboratory}

\author[1]{H.~Ma}
\author[1]{Z.~She}
\author[1]{W.~H.~Zeng}
\author[1]{Z.~Zeng}[orcid=0000-0003-1243-7675]
\cormark[1]
\cortext [cor1]{Corresponding author: zengzhi@mail.tsinghua.edu.cn}
\author[1]{M.~K.~Jing}
\author[1]{Q.~Yue}
\author[1, 2]{J.~P.~Cheng}
\author[1]{J.~L.~Li}
\author[1]{H.~Zhang}

\address[1]{Key Laboratory of Particle and Radiation Imaging (Ministry of Education) and Department of Engineering Physics, Tsinghua University, Beijing 100084}
\address[2]{College of Nuclear Science and Technology, Beijing Normal University, Beijing 100875}

\begin{abstract}
In-situ ${\gamma}$-ray measurements were performed using a portable high purity germanium spectrometer in Hall-C at the second phase of the China Jinping Underground Laboratory (CJPL-II) to characterise the environmental radioactivity background below 3 MeV and provide ambient ${\gamma}$-ray background parameters for next generation of China Dark Matter Experiment (CDEX).
The integral count rate  of the spectrum was 46.8 cps in the energy range of 60 to 2700 keV.
Detection efficiencies of the spectrometer corresponding to concrete walls and surrounding air were obtained from numerical calculation and Monte Carlo simulation, respectively.
The radioactivity concentrations of the walls in the Hall-C were calculated to be ${6.8\pm1.5}$ Bq/kg for ${^{238}}$U, ${5.4\pm0.6}$ Bq/kg for ${^{232}}$Th, ${81.9\pm14.3}$ Bq/kg for ${^{40}}$K.
Based on the measurement results, the expected background rates from these primordial radionuclides of future CDEX experiment were simulated in unit of counts per keV per ton per year (cpkty) for the energy ranges of 2 to 4 keV and around 2 MeV. 
The total background level from primordial radionuclides with decay products in secular equilibrium are ${0.63}$ and ${4.6 \times 10^{-2}}$ cpkty for energy ranges of 2 keV to 4 keV and around 2 MeV, respectively.
\end{abstract}


\maketitle

\section{Introduction}
\paragraph{}
The environmental background in the underground laboratories mainly includes cosmic muons, muon induced particles, ${\gamma}$ rays from rocks and concrete, and neutrons from the (${\alpha}$, n) reaction and spontaneous fission of radionuclides in the surrounding materials.
Since the cosmic rays and cosmogenic radioactivity are drastically decreased with thick rock overburden as shielding \cite{Mei2006}, these underground laboratories \cite{Votano2012, Smith2012, Akerib2004, Ianni2016, Gostilo2020} become favorable sites for rare event searching experiments requiring ultra-low background environment.
Moreover, the muon and neutron background in underground laboratories have been studied comprehensively in literatures \cite{Chazal1998, Robinson2003, Wu2013, Abgrall2017, Hu2016}.

\paragraph{}
The China Jinping Underground Laboratory (CJPL) locates in a traffic tunnel under Jinping Mountain, southwest of China \cite{Cheng2017} with about 2400 m rock overburden vertically. 
CJPL-I has been in normal operation and provided low background environment for two dark matter experiments, CDEX \cite{Yue2014,Jiang2018,Yang2018,Yang2019,Liu2019,She2020} and PandaX \cite{Cui2017} since Dec. 2010. 
The second phase of CJPL (CJPL-II) was proposed to meet the growing space demands from rare event searching experiments. 
The construction of CJPL-II started in 2014 and the cavern excavation was completed in Jun. 2017. 
For better control of background radioactivity of CJPL-II, raw materials of concrete and other construction material were screened and selected by low-background ${\gamma}$-ray high purity germanium (HPGe) spectrometer named GeTHU \cite{Zeng2014} before used.
\paragraph{}
In this paper, in-situ ${\gamma}$-ray HPGe spectrometry was applied to measure the environmental radioactivity of the walls of the Hall-C of the CJPL-II and validate the results of background control measures with GeTHU.
This method is widely used in environmental ${\gamma}$-ray measurements in underground laboratories for its advantage of not requiring sample preparation \cite{Malczewski2012, Malczewski2013, Zeng2014Environmental}. 
This portable HPGe spectrometer was calibrated to obtain the angular response and its detection efficiency of walls of Hall-C based on Beck formula \cite{Beck1972}. 
Since ${^{222}}$Rn and its daughters in the air induce main background for ${\gamma}$-ray measurement, we employed Monte Carlo simulation with GEANT4 \cite{Agostinelli2003} to eliminate their contributions accordingly. 
The next section will describe the setup of in-situ measurements in CJPL-II and the derivation of detection efficiencies in detail. 
The results of measured spectrum and radioactivity concentration will be presented and discussed in the third section. 
The last section will present the expected background level induced by the primordial radionuclide in the concrete walls for the future CDEX experimental facility. 

\section{Experiment and methods}
\subsection{In-situ Measurements}
\begin{figure}
	\centering
	\includegraphics[width = .5\textwidth]{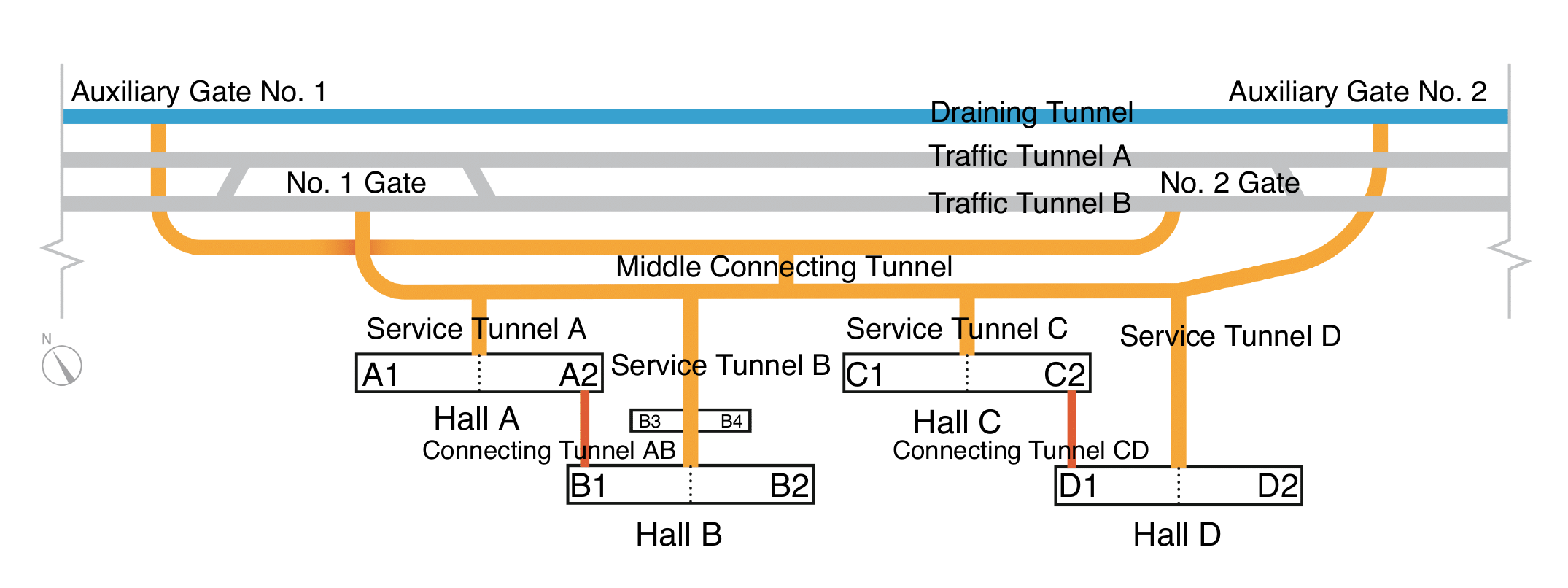}
	\includegraphics[width = .5\textwidth]{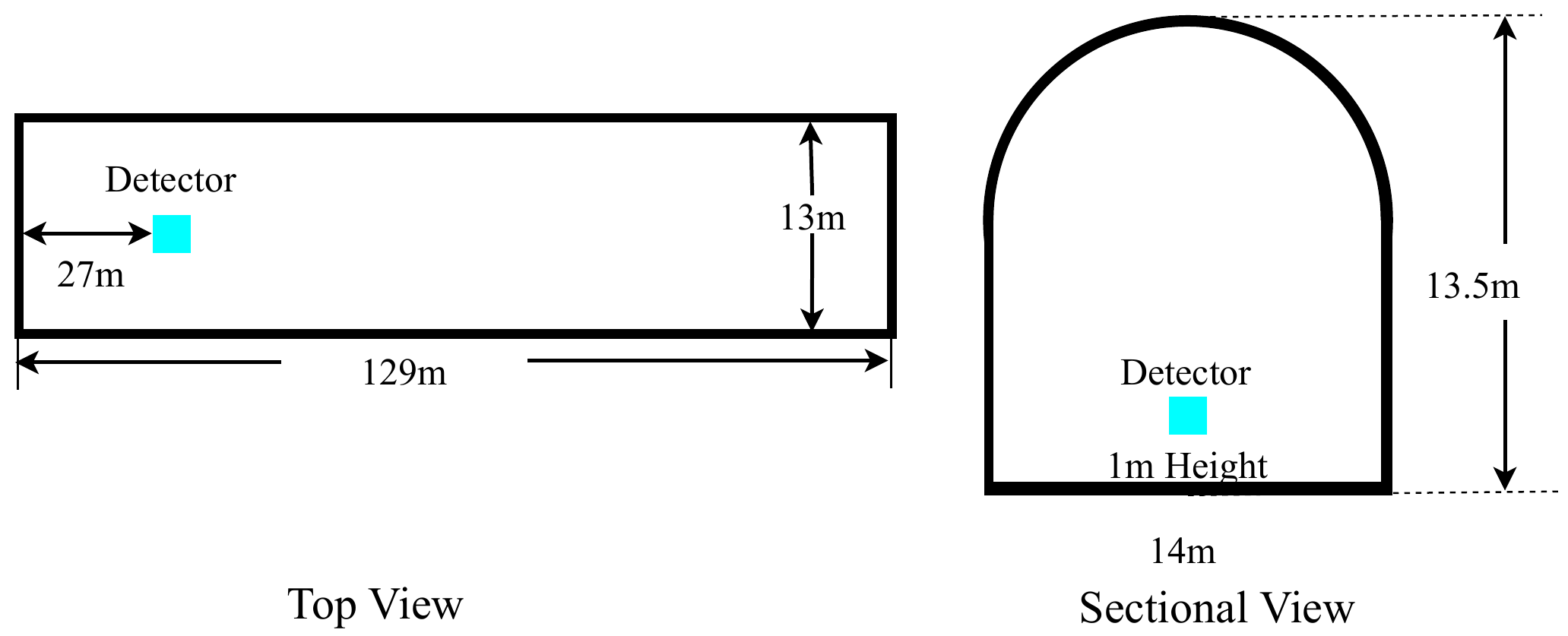}
	\caption{The layout of CJPL-II (top) and the measurement location inside Hall-C (bottom).}
	\label{CJPL-II}
\end{figure}

\paragraph{}
The portable coaxial HPGe detector manufactured by ORTEC was used in the in-situ ${\gamma}$-ray measurements of CJPL-II and its location under measurements in Hall-C is shown in Fig. \ref{CJPL-II}. 
The level of radon, emanated naturally from the concrete and rocks, was measured by an AlphaGUARD radon monitor produced by Saphymo Gmbh which was placed next to the HPGe detector. 
The dimensions of Hall-C were obtained by a laser range finder. 

\paragraph{}
The layout of CJPL-II tunnels and experimental halls are illustrated in Fig. \ref{CJPL-II}.
The CJPL-II has four experimental halls and each hall is divided into two sub-halls with numbers as postfix. 
The China Dark Matter Experiment (CDEX), with scientific goals of searching for light DM, operates germanium detector array in CJPL-I currently. 
Since the future CDEX experiment will be located in the Hall-C \cite{Ma2017} and the dimensions of four experimental halls and their construction materials are almost the same.
In-situ measurements in this work were carried out in Aug. 2019 in the Hall-C of CJPL-II.

\subsection{Detector calibration}
\paragraph{}
The relationship between the rates of ${\gamma}$-ray peaks and the radionuclide activities can be described as Eq. \ref{Beck} and the numerical calculations are applied to determine the detection efficiency for different ${\gamma}$-ray peaks \cite{Beck1972}.
\begin{equation}
\label{Beck}
\frac{N_f}{A} = f(E, \theta) = \frac{N_f}{N_0} \times \frac{N_0}{\phi} \times \frac{\phi}{A} ,
\end{equation}
where ${N_f}$ is rate of the ${\gamma}$-ray peak and ${A}$ is the concentration of the radionuclide ( with the unit of Bq ${\cdot}$ kg${^{-1}}$ for the concrete). 
${f(E,\theta)}$ is the coefficient representing angular response factor multiplying the effective front  area and  ${\theta}$ is the zenith angle between the location of radionuclides and germanium detector.
${E}$ is the energy of ${\gamma}$ ray while ${\theta}$ is zenith angle.
${\phi / A}$ is the ${\gamma}$-ray flux received at detector's location per unit concentration of nuclides.
${N_0 / \phi}$ is the detector response to the incident ${\gamma}$ rays at ${\ang{0}}$ (called the effective front area) while ${N_f / N_0}$ is the detector angular response. 
In this work, the last two were measured in the angular calibration experiment.

\paragraph{}
As shown in Eq. \ref{integration}, we integrate the detection efficiencies with the dimensions of Hall-C. 
i.e., the six surfaces of Hall-C and their geometry boundaries are set as upper or lower bounds in integration. 
To consider the penetrations of ${\gamma}$ rays, the upper limits of integral related to the concrete thickness is set to be the actual thickness (0.2 m). 
The 0.2m-thick concrete will decrease gamma rays from the rock more than an order of magnitude, and the rock has less radioactivities compared with concrete as shown in Table \ref{results}.
In this work, we ignore the penetrating gamma rays from rock.
Following similar procedures described in Ref. \cite{Zeng2014Environmental}, the numerical conversion factors are derived.
\begin{equation}
\label{integration}
\begin{aligned}
    F_{loc} = \frac{r_{cal}^2\rho P}{A_{cal}} \int_{z_{min}}^{z_{max}}\int_{y_{min}}^{y_{max}}\int_{x_{min}}^{x_{max}} f(E, \theta) g_{loc}(x,y,z)  dx dy dz,
\end{aligned}
\end{equation}
where the ${F_{loc}}$ is the numerical conversion factor for different surfaces of the Hall-C.
${r_{cal}}$ and ${A_{cal}}$ are the distance between detector and calibration source, and the calibration source activity, respectively.
${\rho}$ is the concrete density while ${P}$ is the gamma ray intensity.
${(x,y,z)_{min}}$ and ${(x,y,z)_{max}}$ are the lower bound and upper bound of integral, which are related to the different surfaces of the Hall-C. 
${g_{loc}(x,y,z)}$ is the coefficient representing the attenuation effect from concrete and air.
The attenuation coefficients come from the NIST XCOM \cite{XCOM}.
\paragraph{}
The angular correction factors and the effective front areas are determined through angular calibration experiment with a mixed calibration source containing ${^{241}}$Am, ${^{137}}$Cs, ${^{60}}$Co and ${^{152}}$Eu. 
The ${f(E,\theta)}$ surface is fitted by Eq. \ref{fFunction}. 
The relative biases between the fitting curve and the experimental data points are treated as the systematic uncertainties of factors.
\begin{equation}
\begin{aligned}
f(E, \theta)& =  A_{EF} \cdot e^{Q_2(\ln{E})^2+Q_1(\ln{E})+Q_0} \cdot (P_5 \theta^5 \\
& + P_4 \theta^4 + P_3 \theta^3 + P_2 \theta^2 +P_1 \theta +P_0),
\end{aligned}
\label{fFunction}
\end{equation}
where ${A_{EF}}$ represents the effective front area, and ${Q}$ or ${P}$ with numerical subscripts are parameters determined by fitting.
\paragraph{}
Substituting the ${f(E,\theta)}$ in the Eq. \ref{integration}, one gets the numerical conversion factors for certain walls and certain ${\gamma}$-ray energy.
The angular response of detector fluctuates with a relatively small range between ${\ang{0}}$ and ${\ang{150}}$ but decreases rapidly after ${\ang{160}}$, since most ${\gamma}$ rays from the back of detector are shielded by the liquid nitrogen dewar and cold finger. 

\section{Results and discussion}
\subsection{Spectrum analysis}
\begin{figure}[pos=htp]
	\centering
	\includegraphics[width = 0.48\textwidth]{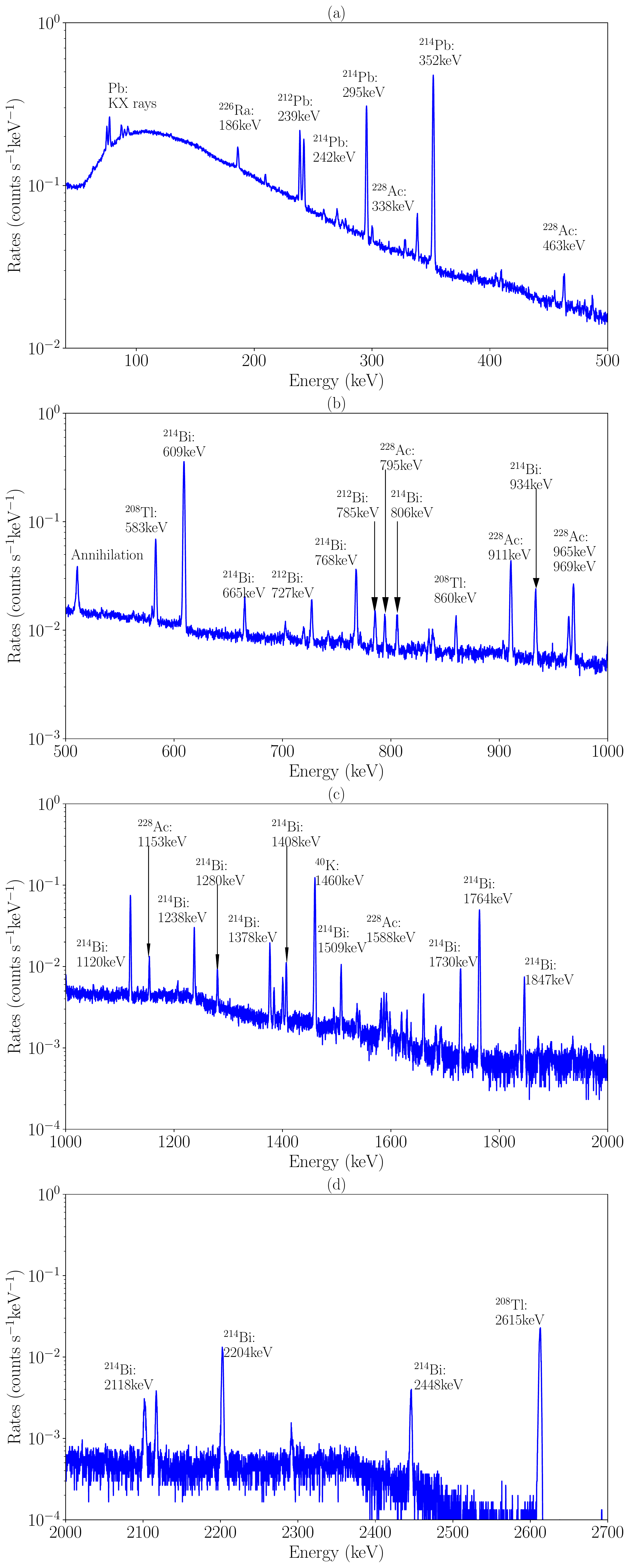}
	\caption{Energy spectra measured in the Hall-C of CJPL-II and labeled with the characteristic ${\gamma}$-ray peaks related to the primordial radionuclides and cosmogenic radionuclides.}
	\label{spectrum}
\end{figure}

\paragraph{}
The energy spectra with the integral rate of 46.8 cps between 60 and 2700 keV was measured in experimental Hall-C with the in-situ ${\gamma}$-ray spectrometer as shown in Fig. \ref{spectrum}. 
The ${\gamma}$-ray peaks from primordial radionuclides are labeled.

\paragraph{}
None artificial radionuclides are found in this spectrum. 
From Fig. \ref{spectrum}(a), the ${\gamma}$-ray peaks (295 keV and 352 keV) are greater than the 186 keV ${\gamma}$-ray peak which is emitted by ${^{226}}$Ra.
However, it reverses when measuring with the concrete samples, so the extra contribution in these ${\gamma}$-ray peaks should come from the ${^{222}}$Rn spread across the Hall-C, which is confirmed by the concentration of ${^{222}}$Rn measured by radon monitor.
The contribution comes from ${^{222}}$Rn must be removed before calculating the concentrations of primordial radionuclides.
\paragraph{}
To extract the integral rates for different ${\gamma}$-ray peaks, the energy spectrum in certain energy range is fitted using a linear continuum plus a single Gaussian peak, the full width half maximum (FWHM) of which is calculated by Eq. \ref{FWHM}. 
The energy resolution of these Gaussian peaks can be fitted with
\begin{equation}
\centering
\label{FWHM}
FWHM(E) = 0.039\times \sqrt{E} + 0.47, 
\end{equation} 
where the ${FWHM}$ is the Full Width Half Maximum of a Gaussian peak while ${E}$ is the energy.
Both of them are in unit of keV.
\paragraph{}
The simulated detection efficiencies of the spectrometer placed in Hall-C are listed in Table \ref{detectionEff}.
Without ventilation system yet in CJPL-II, the Radon concentration was measured to be ${214.5\pm25.7}$ Bq/m${^3}$, contributing a major background to the gamma-ray measurement.
Thus, the detection efficiencies for its characteristic ${\gamma}$-ray peaks are simulated with Geant4.10.5.p01 assuming that the ${^{222}}$Rn together with daughters with short half-lives in equilibrium distributed in the air uniformly shown in Table \ref{detectionEff}. 
Since the space of Hall-C is more than 20000 m${^3}$, a simulation method was adopted to enhance  statistics in this work.
This method, so-called biasing sampling, increases the likelihood that particles have great contributions to improve the Monte Carlo simulation efficiency (see details in Ref. \cite{MCBook}).
\begin{table}
	\centering
    \caption{Simulated detection efficiencies for various ${\gamma}$ rays and the characteristic ${\gamma}$-ray peaks from the experimental wall togenther with${^{222}}$Rn and its daughters in the Hall-C of CJPL-II.}
	\label{detectionEff}
	\begin{tabular}[width=.4\textwidth]{ccc}
		\toprule
		Energy&  Concrete      &  Radon             \\
		keV   &  ${10^{-2}}$(cts/(Bq/kg)) & ${10^{-3}}$(cts/(Bq/m${^3}$)) \\ 
		\midrule
		295.4 &  2.29&  2.5  \\
		351.9 &  4.37&  4.27 \\
		609.3 &  4.76&  3.29 \\
		1120.3&  1.13&  0.61 \\
		583.1 &  3.16&  -    \\
		911.1 &  2.35&  -    \\
		969.1 &  1.36&  -    \\
		1460.8&  0.65&  -    \\
		\bottomrule
	\end{tabular}
\end{table}

\subsection{Ambient radionuclide concentration}
\paragraph{}
The concentrations of primordial radionuclides are calculated following the Eq. \ref{concentration} and listed in Table \ref{results} with measured results of selected concrete and rock samples for comparison. 
The background contributions of detector itself are not subtracted. 
We calculate the average of different gamma peaks related to primordal radionuclides to get their concentrations.
\begin{equation}
\label{concentration}
C = \frac{R - C_{Rn}V\varepsilon}{F_{loc}},
\end{equation}  
where ${C}$ is the concentration of certain radionuclide, and ${R}$ is the integral rate of a specific ${\gamma}$-ray peak. 
${C_{Rn}}$ is the concentration of Rn in the Hall-C. 
${V}$ is the volume of Hall-C while the ${\varepsilon}$ is the simulated radon detection efficiency which is related to the ${\gamma}$-ray energies as shown in Table \ref{detectionEff}.

\paragraph{}
The uncertainties were estimated by error propagation according to Eq. \ref{concentration}.
The uncertainties of peak area include the counting statistics and fitting uncertainties while the uncertainties of radon concentrations were directly provided by AlphaGAURD.
Although the uncertainty of the experimental hall's volume is hard to evaluate due to uneven walls of the hall, the relative uncertainty is assumed to be 10\% conservatively.
The uncertainties of radon detection efficiencies are ignored. 
\begin{table}
	\centering
    \caption{The comparison between in-situ measurements and sample measurements by a germanium spectrometer (the uncertainties include the statistic uncertainties and uncertainties introduced by the Gaussian fitting process). }
	\label{results}
	\begin{tabular}{ccccc}
		\hline
		Nuclides   & Energy &  In-situ &  \multicolumn{2}{c}{Sample measurements} \\
		\ &  (keV) &  measurements& Concrete & Rock \\
		   \         & 	   \ &		(Bq/kg)& (Bq/kg)  &  (Bq/kg) \\
		\midrule
		\multirow{4}{*}{${^{238}}$U}   &  295.4& ${3.8\pm3.6}$  & ${4.9\pm0.07}$& 
		${1.35\pm0.03}$ \\
		\ & 351.9 & ${4.8\pm3.1}$ & ${5.1\pm0.03}$ & ${1.44\pm0.02}$ \\		
		\ &  609.3 & ${7.1\pm2.7}$ &  ${4.1\pm0.03}$ & ${1.23\pm0.02}$ \\
		\ & 1120.3& ${11.5\pm2.6}$ &   ${4.2\pm0.07}$& ${1.27\pm0.05}$       \\ 	
		\multirow{3}{*}{${^{232}}$Th}  & 911.1 & ${5.4\pm1.0}$& ${3.3\pm0.04}$& 
		${1.45\pm0.04}$\\
		\ & 969.1 & ${5.6\pm1.2}$  & ${3.3\pm0.11}$ & ${1.41\pm0.08}$ \\		
		\  & 583.1 & ${5.1\pm0.9}$  &  ${2.9\pm0.03}$& ${1.2\pm0.03}$ \\
		${^{40}}$K                     & 1460.8& ${81.9\pm14.3}$&  ${39.1\pm0.4}$& ${17.3\pm0.3}$       \\
		\bottomrule	
	\end{tabular}
\end{table}

\paragraph{}
The concentration of ${^{232}}$Th and ${^{238}}$U in CJPL-II together with the measured results in other underground laboratories are listed in Table \ref{comparison} for comparison. 
The contamination is less than those of  CJPL-I because of the stricter material screening and selection during the construction.

\paragraph{}
Comparing the in-situ measurements results with sample measurements, the concentration of ${^{40}}$K differs significantly from that of concrete sample.  
The in-situ measurements actually assess the average concentrations across the hall while the concrete samples are just pieces of the surrounding concrete walls. 
The measured concentrations for various concrete samples differ greatly especially for ${^{40}}$K (4.0 - 157.0 Bq/kg). 
\begin{table}
	\centering
	\caption{The concentrations of primordial radionuclides in different underground laboratories.}
	\label{comparison}
	\begin{tabular}{ccccc}
		\toprule
		Underground Lab& ${^{238}}$U& ${^{232}}$Th& ${^{40}}$K	\\
		               &     (Bq/kg)&   (Bq/kg)   &    (Bq/kg)  \\ 
		\midrule
		Gran Sasso \cite{Malczewski2013} &  ${9.5\pm0.3}$& ${3.7\pm0.2}$& ${70\pm2}$   \\
		Modane \cite{Malczewski2012}     & ${22.8\pm0.7}$& ${6.7\pm0.2}$& ${91\pm3}$   \\
		Boulby \cite{Malczewski2013}    &  ${7.1\pm0.2}$& ${3.9\pm0.1}$& ${120\pm2}$  \\
		Sanford \cite{Akerib2020} &  ${29\pm15}$& ${13\pm3}$& ${220\pm60}$ \\
		CJPL-I \cite{Zeng2014Environmental}   &          18.0 &          7.6 &	     36.7   \\
		CJPL-II   &   \multirow{2}{*}{${6.8\pm1.5}$}&   \multirow{2}{*}{${5.4\pm0.6}$}&   \multirow{2}{*}{${81.9\pm14.4}$}  \\
		(this work) & \ & \ & \ \\
		\bottomrule 		
	\end{tabular}
\end{table}

\section{Background assessment for CDEX-100}
\begin{figure}
	\centering
	\includegraphics[width=\linewidth]{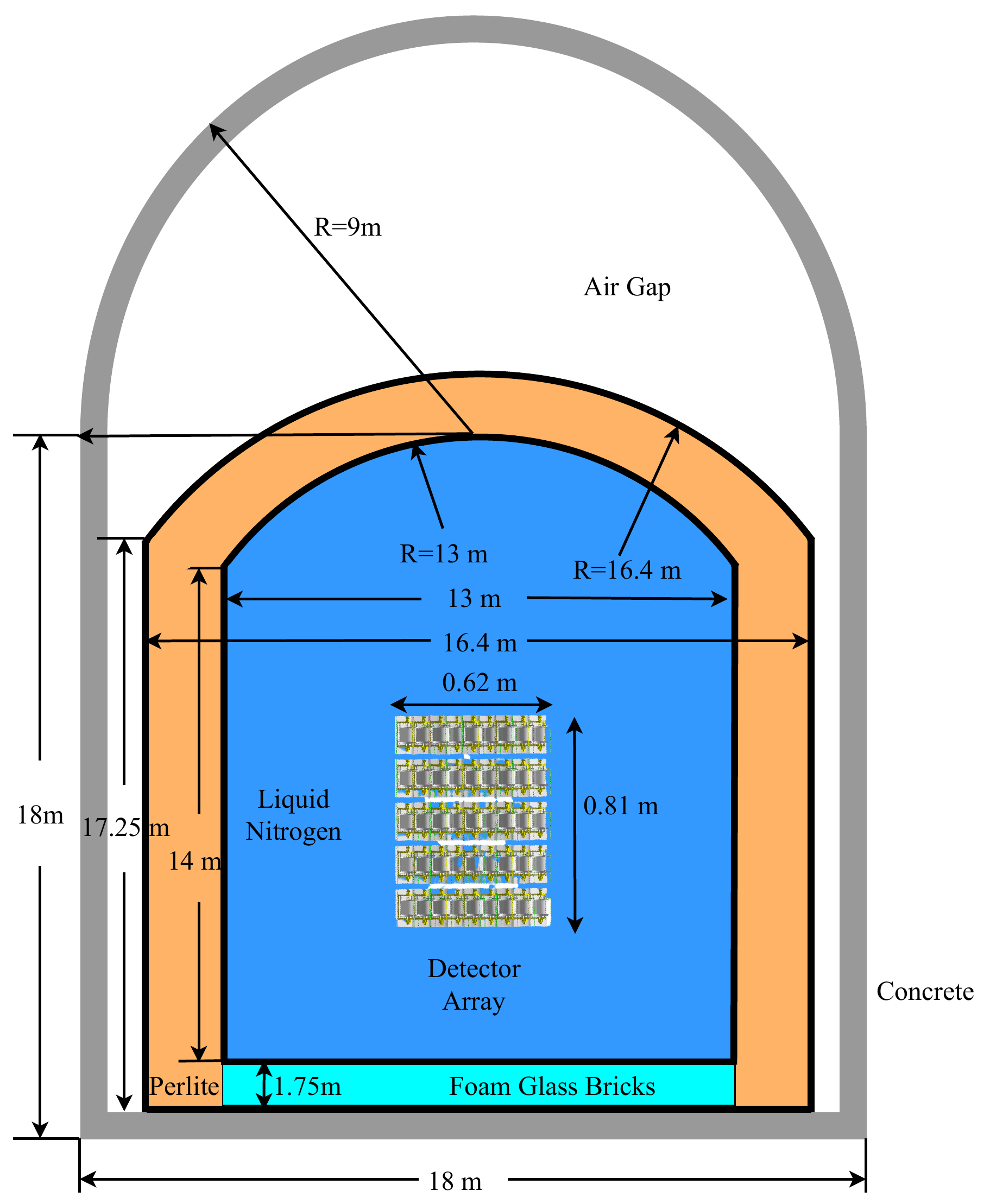}
    \caption{The schematic diagram of the CDEX-100 experimental setup (not to scale). The concrete (0.2 m), perlite (1.7 m), foam glass bricks (1.75 m) as heat insulator and liquid nitrogen are drawn in gray, orange, cyan and blue, respectively. The steel tank and cryostat are shown with black solid lines. The air gap between the tank and the concrete wall is included in simulation. All of these rectangles in this figure are the section views of cylinders.}
	\label{simulation}
\end{figure}

\paragraph{}
The next phase of CDEX experiment named CDEX-100 will operate about 100 kg germanium detector array for enlarging the target mass and study background characteristics in the liquid nitrogen environment \cite{Ma2019}. 
The CDEX germanium detector array will be installed in a cryotank, whose volume is about 1725 m${^3}$, immersed in the LN${_2}$ at Hall-C of CJPL-II, and the 6.5 m-thick LN${_2}$ will act as the passive shielding from ambient radioactivity.  
\paragraph{}
Importing the measured concentrations of primordial radionuclides, the background simulation was conducted with a dedicated Monte Carlo framework called Simulation and Analysis of Germanium Experiment (SAGE) based on Geant4.10.5.p01 \cite{Agostinelli2003} for CDEX-100 experiment. 
The radionuclides are assumed to be uniformly distributed in the concrete wall close to the cryotank as shown in the Fig. \ref{simulation}. 
The gap between the inner and outer tanks is filled with the perlite and foam glass bricks as heat insulator.   

\paragraph{}
Thanks to the 6.5 m thick liquid nitrogen, gamma rays from primordial radionuclides could hardly penetrate.
The uniformly distributed source for certain radionuclide at the inner surface of the concrete are assumed in the simulation rather than a volume source actually to get a higher simulation efficiency.
The emitting efficiency of gamma-rays in the concrete wall was simulated separately and used to determine the surface source activity.
The aforementioned biasing techniques were implemented to reduce statistic uncertainty.

\paragraph{} 
The regions of interests (ROIs) in CDEX-100 are the energy ranges from 2 to 4 keV for light dark matter detection and around 2 MeV (2014 to 2064 keV for ${^{76}}$Ge neutrinoless double beta decay).
The background rates of the two ROIs are simulated for all germanium detectors without any vetos or pulse shape discrimination.
For ${^{238}}$U, the background rates are ${0.25}$ and ${1.5 \times 10^{-2}}$ cpkty for these energy ranges, while ${0.38}$ and ${3.1 \times 10^{-2}}$ cpkty for ${^{232}}$Th. 
Due to the lower Q value, ${^{40}}$K will only contribute ${5.3 \times 10^{-4}}$ cpkty to the range of 2 to 4 keV.

\section{Conclusion}
\paragraph{}
CJPL-II is a deep underground laboratory designed for future large-scale rare event searching physics experiments.
We have applied in-situ ${\gamma}$-ray HPGe spectrometry to measure the ${\gamma}$-ray background in the Hall-C of CJPL-II.
The numerical calculation based on the calibration of angular responses and Monte Carlo simulation were used to obtain detection efficiencies of ${\gamma}$ rays from walls and surrounding air.
After reduction of radon contribution, the radioactivity concentrations in the concrete are characterized as ${6.8\pm1.5}$ Bq/kg for ${^{238}}$U, ${5.4\pm0.6}$ Bq/kg for ${^{232}}$Th and ${81.9\pm14.4}$ Bq/kg for ${^{40}}$K, competitive with other underground laboratories.
The background level from primordial radionuclides is simulated for the future CDEX-100 experiment. 
The background rates from primordial radionuclides are ${0.63}$ and ${4.6 \times 10^{-2}}$ cpkty in the ROIs of 2 to 4 keV and 2014 to 2064 keV, respectively. Other intrinsic backgrounds from the experimental setup will be studied in future work to get a whole background map in CDEX-100 experiment.

\section*{Acknowledgement}
We gratefully acknowledge support from National Key R\&D Program of China (No. 2017YFA0402201), National Natural Science Foundation of China (No. 11725522 \&
11675088), and Tsinghua University Initiative Scientific Research Program (No.20197050007). 
\bibliographystyle{unsrt}
\bibliography{./CJPL}
\end{document}